# PAPER SCIENCE

A. Müller

*Institut für die Pädagogik der Naturwissenschaften Olshausenstr. 62,*

*24098 Kiel (Germany)*

[mueller.andreas@ipn.uni-kiel.de](mailto:mueller.andreas@ipn.uni-kiel.de)

Paper has a lot of interesting properties with which quite a lot of standard topics of science education can be turned into hands-on activities. Among others, experiments on elasticity, capillarity, feedback oscillations, flight, friction, perception and thermal expansion will be shown.

## 1. Introduction. An Experiment on Negative Feedback

The experiments presented below are on different levels, starting from primary school up to open questions of science. An important factor determining the level is whether qualitative, semiquantitative or quantitative understanding is required. The same experiment can be suitable for school children, when focussing on a qualitative explanation, and for university students, when aiming at a quantitative theory (an example for this will be given in last section). Several of the experiments have been tested in one or two levels of science education; this will be indicated.

The experiments can also be situated in different classroom contexts, e.g. as group work, home work and open ended investigations. Another dimension for context are applications of the investigated phenomenon to everyday life (lifeworld contexts) or in other sciences (cross-disciplinary contexts).

A general remark on hands-on-experiments and other science activities is in order. There are very many ideas like the collection presented below, showing the same variety and flexibility in level and context – in principle an advantageous situation for science teaching. This advantage, however, suffers from a practical difficulty: An in-service-teacher (as well as a lecturer at the university – there is little difference in that respect) will often not have the time to look through these sources and to find something which is appropriate to his specific classroom purposes. A tool is necessary, which allows science educators to select science activities in accord with their goals, curriculum and context. Such a tool is offered by modern technology would be a computerised data base which allows a specific and direct selection of science activities according to given topical, educational and practical criteria. I have undertaken first steps to develop such a database: there is a collection of several hundred examples, a tentative structure of the database entries (see Fig. 1), and support for the programming work (which will begin very soon). Whoever has similar interests is strongly encouraged to contact me – I am very interested in cooperation for this project!

The first example for a Paper Physics experiment is at the same time an example of a database entry in the tentative format of Fig. 1 (some more examples will follow in the next sections). For lack of space, not all items (fields) of the structure are considered. For the same reason, only a few of the examples in the following sections will be given in this detailed form.

**Experiment 1: Negative Feedback / Paper Flute**

**Material, equipment:**

A square sheet of paper, of size 10-20 cm, best obtained as a photocopy of Fig. 12 (see appendix); a round stick about 10 cm long and 0.5 cm in diameter (e. g. a pencil), sticky tape; a pair of scissors.

If you do not start with a photocopy, a ruler is necessary in order to draw the bending and cutting lines with the necessary accuracy.

**Content Features**

Description: Make the cuts as shown in the drawing. Then form a paper tube by rolling the sheet around the round stick; the triangular flap should remain flat. Prevent the tube from unrolling with the sticky tape, and prevent the mouth piece of the flute (opposite to the triangular flap) from being softened by saliva by another piece of sticky tape. Then bend the flap slightly towards the opening of the flute.

On sucking (!) air through the flute, it gives a buzzing sound (the children use to find quite a specific name for it). It is possible that you have to experiment a little with suction strength to make the flute work

Explanation: Through the suction of the air, the flap is pulled towards the opening of the paper tube, and eventually closes it. The air can then no longer flow, and the flap bends back towards its initial position. With the increasing width of the opening, the air suction sets in again, and a new cycle begins.

This is an example of a process (the air flow) which hampers or even stops itself (the flap closes the opening). Such processes are very widespread both in nature and technology, and their working principle is called *negative feedback*.

Applications: Negative feedback oscillators are at work in other fields, e.g. bow instruments, the self-interruption circuitry in an electrical bell or a car winker.

**Educational Features**

Questions, Tasks:
Try to get a sound from the flute (first without telling that one has to suck). Show and explain the flute to your friends and family. Give a *written* description and explanation of the experiment.
Observe how the sound changes with: (i) the diameter of the flute; (ii) the width of the paper bridge connecting the flap with the body of the flute; (iii) the strength of your suction; (iv) the stiffness of the paper. Explain the dependencies you found.

Learning goals:
Role of accuracy in manufacturing and observation (see practical hints below).
Experience with general, system independent principles overarching different areas of science.

**Practical aspects (see also material/equipment)**

Age: Starting from 8-9.
Level: intermediate - high (depending on age)
Duration: 1-2 h (depending on age and on length of discussion)
Practical hints: The flute does not work if the following steps are not carried out quite accurately: (i) the cuts for the flap must be perpendicular to the symmetry axis of the paper; (ii) the rolling of the paper must be parallel to the symmetry axis of the paper.
The paper should not be rolled too tightly around the pencil, which otherwise is difficult to pull out.
Security: Usage of a pairs of scissors.
Test: Primary school (mixed class of 3rd and 4th graders), teacher education

## 2. Experiments on Mechanics

### 2.1 Elementary Mechanics

**Experiment 2: Statics (Stress Concentration)/ Paper Strip**

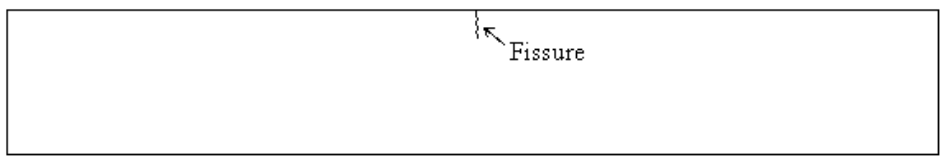

Take two paper strips like in Fig. 2, one with the little fissure in the centre, the other without; pull at the ends of the strips; the one with the fissures rips much more easily than the one without the fissure. A first factor which comes to ones mind is *the proportionality of rupture strength and cross section*. It is indeed true that the maximal rupture load for a strip of paper is proportional to its width (all other conditions fixed), but this cannot explain the extend to which rupture is facilitated by a fissure.

A second simple experiment can prove this: A paper strip like in Fig. 3 will rip at the fissure, even though the width is larger there than at the left end.

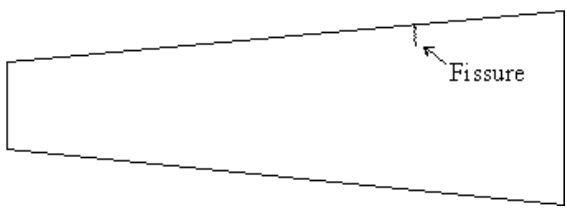

As second and decisive factor is the *stress concentration* around a fissure, which is described by

$$\sigma \propto r^{-1/2},$$

i.e. the stress $\sigma$ diverges when the radius of curvature $r$ of a fissure tends towards zero. Stress concentration has many practical consequences, e.g. it is the reason why glass "cutting" works.

**Experiment 3: Statics (Moment of Inertia) / Paper Ring**

Bend the paper strip of Fig. 4 to a ring and close it by sticking the cuts into each other. You obtain a ring which is able to carry a glass of water or a comparable weight. The experiment shows that a very weak material (here: paper) can give a very stable structure, when in the right shape (here: a ring or a tube). Scientifically spoken, the buckling load of a rod is proportional to its moment of inertia $I$, and with a given amount of material, the ring has maximal $I$. This is known both to architects and mother nature: television towers and plant spears often have roughly the shape of a tube.

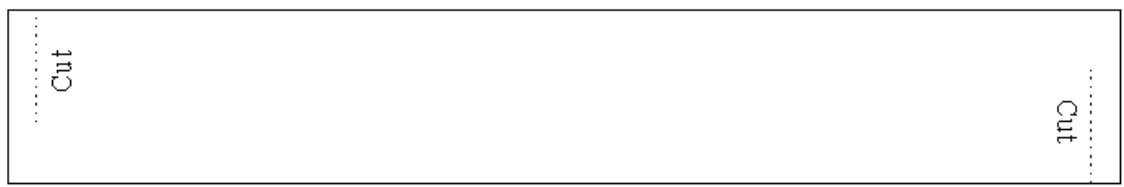

**Fig. 4: Paper Ring**

### 2.2 Aerodynamics and Physics of Flight

This area offers almost endless possibilities for Paper physics, viz. building and investigating paper airplanes. Many nice examples of this type of physics toys are well-known, and there is no point to enlarge the existing collections (see Ref. 1). Rather, some selected examples will be presented, where a single or a few physical concepts beyond mere air lift clearly appear.

**Experiment 4: Air Resistance / Paper Parachute**

Prepare a sheet of paper as described in Fig. 13. Tie four threads of about 40 centimetres length at the holes, then tie together their loose ends in a single knot (make sure that all threads have equal lengths between the holes and this knot). Then fasten a metal paper clip at the knot (you may make a second knot to fix the clip) and bend one of its ends to a little hook where a weight can be attached (bolts are quite appropriate as weights).

The experiment provides a simple, though appealing way of experiencing and experimenting with air resistance; several semiquantitative variations are possible, e.g. measuring the falling time as function of the attached weight and of the size (surface) of the parachute.

**Experiment 5: Aerodynamics & Top Physics / Paper Boomerang**

In order to make the Paper boomerang work, you have to enlarge Fig. 5 so that the length of a wing is 4-5 centimetres. Copy the black boomerang-like contour on cardboard (of about the thickness of a postcard). Twist the right wing a little upwards, the left wing a little downwards. Put the boomerang on your left palm (held horizontally) as shown. To make the boomerang go, flick it with one of the fingers of your right hand. (For left-handers, left and right in the foregoing description have to be interchanged)

The theory of the boomerang is involved, but readable introductions do exist, see Ref. 2. I have the impression, however, the returning of our little Paper boomerang resembles more that of a Frisbee disk, thrown obliquely upwards. Whether we are dealing with a boomerang or Frisbee return mechanism or a mixture of both seems to me an open question, left to your own investigations.

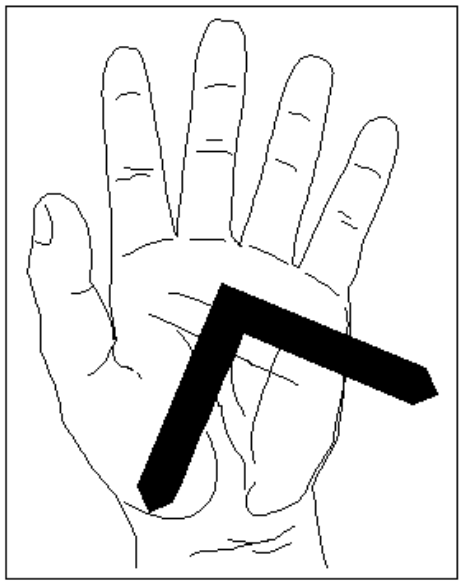

Fig. 5: Paper Boomerang

## 3. Experiments on Thermodynamics and Physical Chemistry

**Experiment 6: Thermal Expansion / BiPaper**

Bimetallic temperature switches and similar devices are a standard application for the topic of thermal expansion in school. Fig. 6 shows how the same experiment can be done with BiPaper (e.g. the paper from a cigarette box) instead of bimetal. Instead of a cigarette tip as heat source you can also keep the BiPaper with tweezers and heat it with a candle flame.

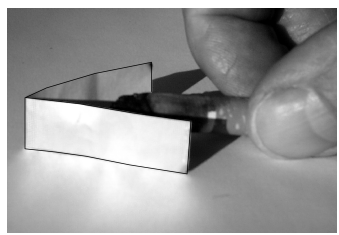

**Fig. 6: Bi Paper**

**Experiment 7: Capillarity / Paper Flower** (This experiment is given in the format of Fig. 1)**.**

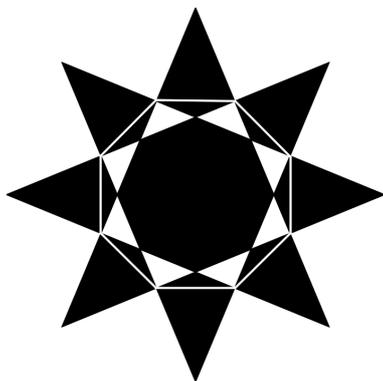

**Fig. 7: Paper flower**

Material, equipment: Ink blotting paper or ordinary writing paper, a glass or little bowl with water, a pair of scissors.

**Content Features**

Description: Cut a blossom shaped or star shaped piece as in Fig. 7 from the ink blotting paper and fold it inwards along the white lines; the fold should be made quite sharp (e.g. by rubbing along it with your fingernails). Then put it on the water surface. Within less than a minute (sometimes within seconds, depending on the kind of paper you are using) the blossom unfolds.

Explanation: The capillaries in the paper are straightened out by the capillary pressure of the water "soaked in". You can show this straightening effect by blowing in a tubular plastic wrap or bag (e.g. the wrapping of some salami sausages).

Applications: The capillary force is indeed the reason for a part of plant movements such as the opening of blossoms (another contributing force here is the osmotic pressure). The capillary force in wood is also used to quarry stones in ancient quarries and stonemasonry.

Interdisciplinary links: Physics-chemistry (surface Science)-biology (plant physiology)

**Educational Features**

Level: Easy
Type: Group work, home work
Questions, Tasks: Show and explain the experiment to your friends and family. Give a *written* description and explanation of the experiment. Investigate different types of paper: Is there a relation between how "soaky" it is and how fast the effect occurs? Can you give other examples where the capillary force it at work?

**Practical aspects (see also material/equipment)**

Age: Starting from 8-9
Duration: 1-2 hours, depending on how many of the related experiments you want to be made.
Security: Use of a pair of scissors
Test: Primary school (mixed class of 3rd and 4th graders), teacher education

## 4. Experiments on Optics and Visual Perception

As an outlook on a wide field of applications of Paper Science I want to point out at a particularly interesting area of biophysics and psychophysics, viz. optics and visual perception. One experiment in this field is described below another famous experiment is the neuronal image inversion (see e.g. Ref. **3**), and numerous further experiments can be made on optical illusions with stimuli on printed paper.

**Experiment 8:** Pinhole Optics / Paper "Glasses" (This experiment was tested in a special course on perception for gifted pupils (12th and 13th grade of German gymnasium) and in teacher education.)

Fig. 9 shows the probably cheapest glasses you can imagine: you can make it directly from a copy on cardboard and making the holes with a nail, or you can copy it at a transparency.

For the experiment, ask a long-sighted person to approach just below the minimal distance where she can still read some text (because of their size, newspaper headings are most suitable for classroom demonstrations). Now let her put on the glasses and look again at the text. Surprise! The text is well readable now, and this will produce not only a strong effect on the person herself, but also on the spectators watching her.

The explanation is, of course, the well-known pinhole effect, see Fig. 8: if the peripheral part of the divergent light bundle emanating from a source point is screened by a diaphragm, the image disk (central part of light bundle) will be quite small, even for a long-sighted person (a similar explanation holds for a short-sighted person). The price for this is a rather poor comfort for the user of the pinhole optics, due to a reduction of light intensity by a factor $2\pi r^2/d^2$ and a reduction of the field of sight by (roughly) a factor $r^2/a^2$ (where $r$ is the radius and $d$ the distance of the holes, and $a$ is the distance of the holes and the image point on the retina; the ratio for the field of sight with hole ($2\pi r^2/a^2$) and without hole ($2\pi$) is only approximate, disregarding for the field of sight through the holes the refraction in the eye and overestimating the field of sight of the naked eye as $2\pi$). All this is not new, but what might be new is the idea to improve the single pinhole optics by arranging them in an array (at least I did not see this idea before). This improvement has a limit, where the image disks of the different holes start to overlap.

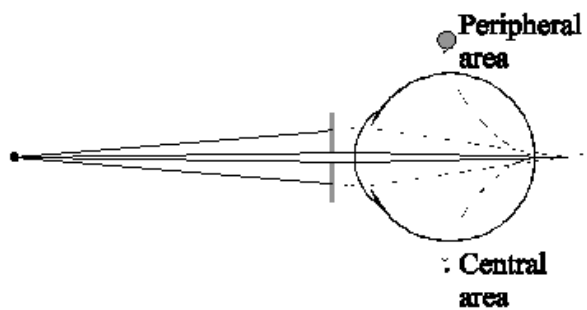

Fig. 8: Pinhole Optics (for a long-sighted person)

A nice exercise (and, as in several preceding experiments, a step to a more difficult level of physics) is to determine the optimal hole size for given properties of the eye, i.e. the largest size where the images do not yet overlap (overlapping means blurring). Hint: These properties include, of course, the number of diopters. What else? Note that the border of image disk you see is actually a image of your *pupil* (!). This can be verified by suddenly varying the brightness conditions, thus changing the pupil size and disk size simultaneously. The optimal hole size thus depends on the pupil size, too.

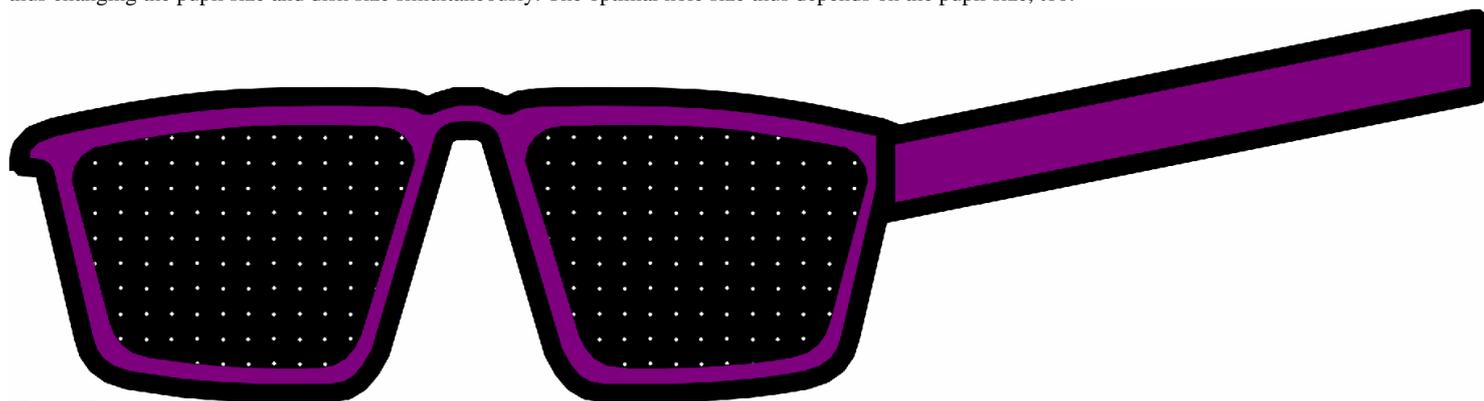

**Fig. 9: Paper glasses**

# 5. Conclusion and Outlook. An experiment on positive feedback

The experiments described here should be sufficient to achieve the main purpose of this contribution, viz. to stimulate the phantasy of science educators by showing that very common (and thus cheap!) materials can be used for a whole collection of science investigations which have several educationally desirable features at a time:

- Activity oriented (toy character, self-built devices)
- Connections to everyday life (use of a common material, application to everyday phenomena )
- Applications to real-world problems (importance of stability, friction etc. in engineering)
- Cross-disciplinary links (biophysics, physical chemistry, material science)
- Motivation to *think* about scientific phenomena (resulting from all of the above-mentioned factors)

The examples shown range in level from a playful investigation in primary school to a challenging task at university and they cover topics from mechanics, acoustics, optics, thermal physics, including some points of contact to physical chemistry and material science.

As a further outlook a last experiment on feedback will be presented, thus making a full turn back to the first experiment, and showing again how even very general principle like feedback, overarching various branches of sciences, can be demonstrated with Paper Science. From here, numerous other applications and transfer exercises for feedback in other contexts can be undertaken.

**Experiment 9: Paper Friction**

Material, equipment: Two pocketbooks of about the same size (at least 150 pages and the rougher the paper the better).

**Content Features:**

Areas and subareas of science: Mechanics
Curriculum links: Friction
Description: Interleave the pages of the two books in a way that the pages of the one book overlap with those of the other by at least 10 cm (card players know a similar way of shuffling). The result is displayed in Fig. 10: In the middle a pile of pages as thick as the two books together, on the sides the backs of the two individual books.

Now hold the two books some centimetres away from the thicker part in the middle and try to pull them apart – even with all your force you will not succeed!

Explanation: The books are held together by the friction force, the size of which is quite surprising, as the pulling force of your arms easily attains 100-200 *N*. Where does this large friction force come from?

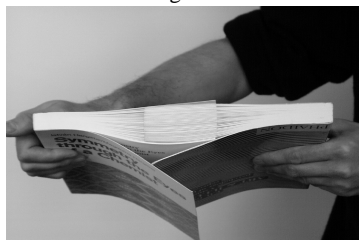

**Fig. 10: Paper friction**

A first and obvious idea is that in order to get a total friction force $F_T$ between the two books you have to multiply the friction force $F_F$ holding back a single leaf by the number $S$ of leaves. To make the following estimation as simple as possible, it will be assumed that the two books are identical and that each individual leave will be interleaved (i.e. $S = ½ P$, where $P$ is the number of pages in the book), even though practically one will rather interleave *groups* of leaves. Despite of this fact, I will speak about "leaves" even when in reality there are "groups of leaves", and the quantitative error (due to $S < ½ P$) can be accounted for in the final result.

*It is of course correct that the friction force is proportional to the number of leaves, and one could be tempted to be satisfied with this explanation. However, the argument does not stand an order of magnitude estimate of the forces involved. We have:*

$$F_T = S \cdot F_F \geq 100\ N$$
$$F_F = 2\mu\ F_N\ ;\ \mu = 0.3 – 0.5$$

Eq. 1

where $F_F$ has been expressed by the normal force $F_N$ and the friction coefficient μ for paper on paper (the approximate value given for μ can he easily found in a separate experiment, and the factor 2 in the second line is due to the fact that each leave has two sides).

It follows from the above equation that $F_N$ has to be at least some Newtons. Where does this normal force come from? It can not be the gravitational force, because the effect also occurs if the books are held vertically, i.e. if there is no component of the gravitational force normal to the leaves at all. We therefore must suspect another factor at work, different from and less obvious than mere multiplication with the number of leaves.

A more thoughtful inspection of the situation reveals that the leaves are bent outside by an angle φ , see Fig. 11. This is because the paper pile in the middle has the thickness of the two books together, whereas on the sides it has only the thickness of each individual book. The resulting geometry then makes obvious that a normal force in the required direction arises as normal component of the pulling force along each leave.

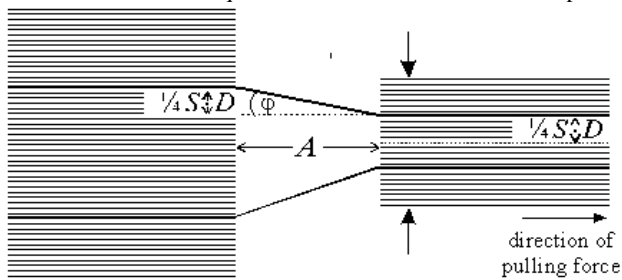

**Fig. 11: Geometry for the explanation of the self-blocking effect** (see Eq.2 through Eq. 6). The figure shows the right side of Fig. 10. The bold vertical arrows indicate the points book is held, i.e. where the holding force is applied.

To make sure that we have found now a satisfactory explanation we estimate the resulting friction force as follows. The normal force exerted by a single leave is proportional to its share of the total pulling force and given by

$$\frac{1}{S} \sin\varphi \cdot F_z \quad . \qquad \text{Eq. 2}$$

The angle φ and thus this contribution are not the same for all leaves, but rather depend on the position in the stack. To get an estimate, we can consider the leaf with running number *S/4* (in the pile to the right, counting from the horizontal symmetry axis, see Fig. 11) as a typical example. The angle is then given by

$$\tan\varphi = \frac{SD/4}{A} \quad . \qquad \text{Eq. 3}$$

Note that for small angles $\tan\varphi \approx \varphi \approx \sin\varphi$.

The normal force $F_N$ on a given leave is the sum of the contributions of all leaves pushing on it. This sum can be estimated as follows: There are S/2 leaves pushing on the one considered, and they are roughly exerting the same normal force as given by Eq. 2. Thus

$$F_N \approx \frac{S}{2} \cdot \frac{1}{S} \sin\varphi \, F_z \approx \frac{1}{8} S \frac{D}{A} F_z \quad , \qquad \text{Eq. 4}$$

(where Eq. 3 and the approximation below it was used for φ). The true value of $F_N$ is rather larger than stated by Eq. 4, because the angle φ and thus the normal force components arising from the leaves further outside are larger than in Eq. 2. It then follows from Eq. 4 and Eq. 1 that

$$F_T \approx \frac{1}{4} \mu S^2 \frac{D}{A} F_z \quad . \qquad \text{Eq. 5}$$

This is the final result and shows two remarkable features: First, the total friction force $F_T$ is proportional to the pulling force, i.e. the stronger you pull the books apart, the stronger they hold together. Second, the holding force will be larger then the pulling force if the constant of proportionality is larger than one:

$$\frac{1}{4} \mu S^2 \frac{D}{A} \geq 1 \quad . \qquad \text{Eq. 6}$$

This condition can be easily fulfilled with increasing *S,* and if it is fulfilled you may pull as strong as you want and the books will not get apart. This phenomenon is precisely what we observed in the experiment. It is called self-blocking.

As a last step in our consideration we have to account for the approximation explained above Eq. 1: the fact that groups of leaves instead of all individual leaves are interleaved makes the total friction force in Eq. 5 and the parameter combination in Eq. 6 smaller by roughly a factor 3 or 5; this does not change, however, the main argument presented here.

**Educational Features:**

**Type:** Semiquantitative
**Learning goals:** Critical attitude towards superficially plausible explanations. Importance of order-of-magnitude estimates and semiquantitative understanding.

**Practical aspects (see also material/equipment):**

**Difficulty:** high
**Practical hints:** You should make sure that you interleave as many leaves as possible. With a book of 150 pages, however, it does not matter when there are groups of several pages instead of separate ones interleaving.
**Test**: In teacher education.

# Acknowledgements:

Lots of people have given me ideas on Paper Science experiments. In particular, Experiment 2 about stress concentration was brought to my attention by Prof. A.I. Pesin, Charkov, right after my workshop at the GIREP conference, and the Paper Boomerang (Experiment 5) was shown to me by H.-F. Schleichermacher, Lübeck, an expert for boomerangs.

**Appendix: BluePrints for Paper Science**

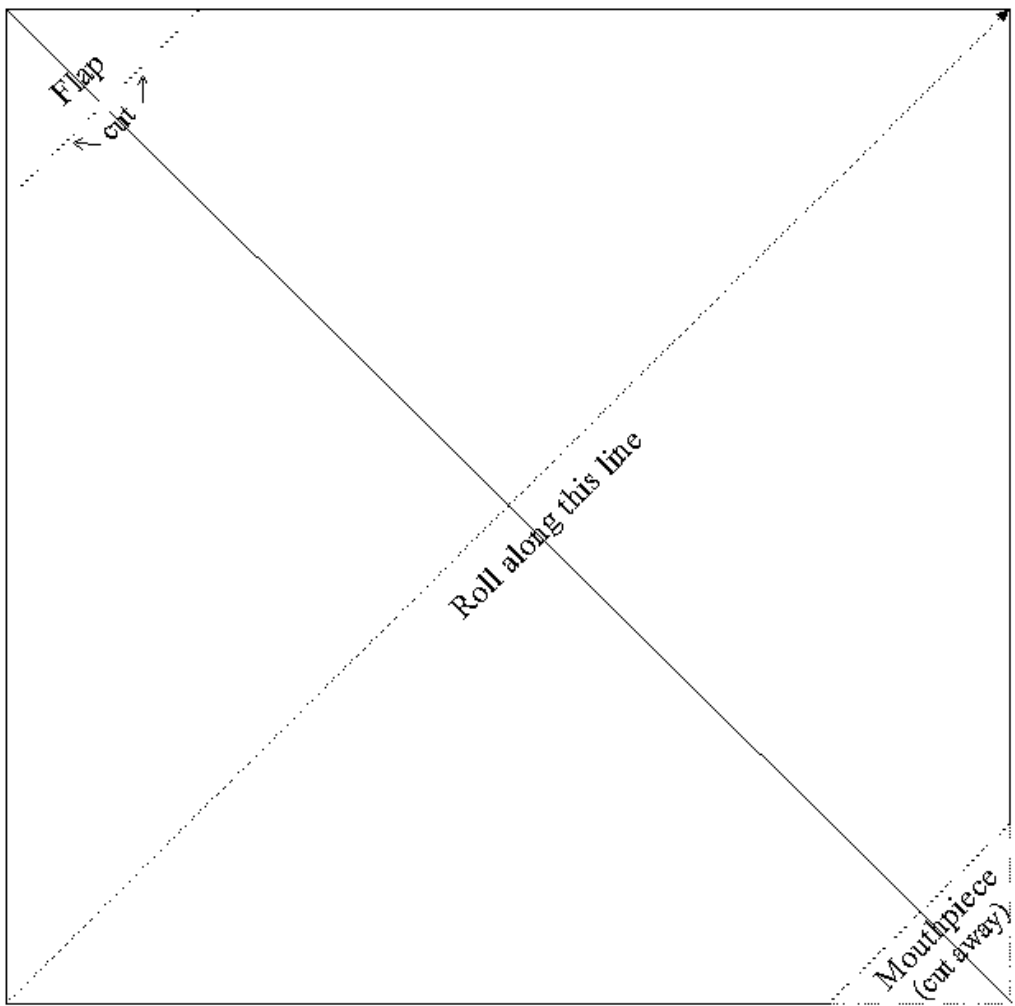

**Fig. 12: Paper Flute (enlarge to 10-20 cm; make sure that the cuts in the upper left corner stop about 2 or 3 millimetres from the diagonal)**

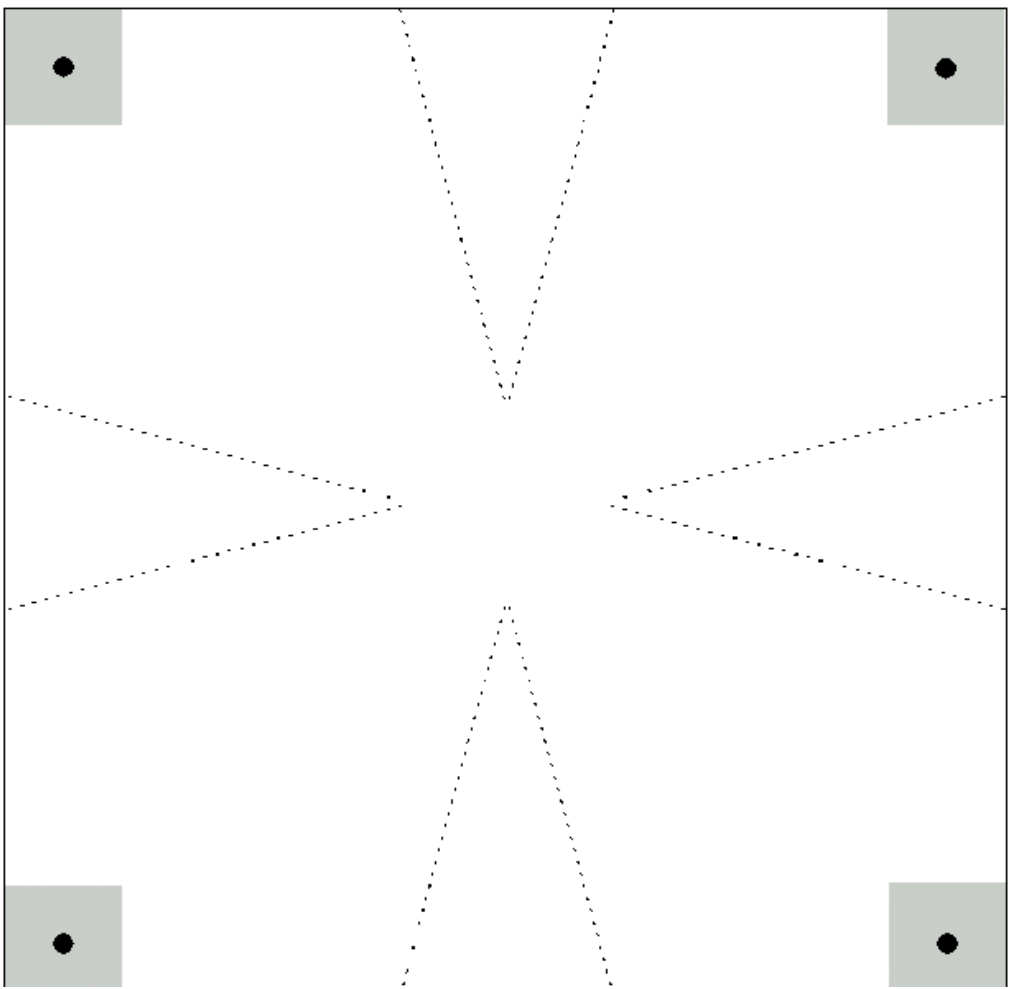

**Fig. 13: Paper Parachute (enlarge to 15-20 cm)**

— 1st  cut along solid lines,
..... 2nd  cut along dotted lines

gray:   Strengthen paper first in these areas with sticky tape
black dots: Holes, e.g. by a hole puncher

Erratum:
The number of leaves pushing on the one considered in experiment 9 is S/4, not S/2 (see eq. 4 and above).
This reduces the factor 1/4 in eq.6 to 1/8.